\documentclass[]{mn2e}
\usepackage{times}

\title{A New Wolf-Rayet Star and its Ring Nebula: PCG\,11}
\author [Martin Cohen, Q.A. Parker, Anne J. Green]
{Martin Cohen$^{1}$, Quentin A. Parker$^{2,3}$, Anne J. Green$^{4}$ \\
$^{1}$Radio Astronomy Laboratory, University of California,
    Berkeley, CA 94720\\
$^{2}$Department of Physics, Macquarie University, Sydney, NSW 2109 Australia\\
$^{3}$Anglo-Australian Observatory, PO Box 296, Epping, NSW 2121, Australia\\
$^{4}$School of Physics, University of Sydney, NSW 2006, Australia\\}

\date{Accepted
      Received
      in original form        }

\begin{document}

\maketitle

\begin{abstract}
In a search for new Galactic planetary nebulae from our systematic scans of the
Anglo-Australian Observatory/United Kingdom Schmidt Telescope (AAO/UKST)
H$\alpha$ survey of the Southern Galactic Plane, we have identified a Population~I Wolf-Rayet
star of type WN7h associated with an unusual ring nebula that has 
a fractured rim.  We present imagery in H$\alpha$, the 843-MHz continuum from the Molonglo 
Observatory Synthesis Telescope (MOST), the mid-infrared from the Midcourse Space Experiment 
(MSX), and confirmatory optical spectroscopy of the character of the nebula and of its 
central star.  The inner edge of the H$\alpha$ shell shows gravitational instabilities with 
a well-defined wavelength around its complete circumference.
\end{abstract}


\section{INTRODUCTION}
The high-resolution AAO/UKST H$\alpha$ Survey of the Southern Galactic plane (Parker \& Phillipps 1998)
is providing an unprecedented new source of Galactic planetary nebulae (PNe) thanks to the survey's
powerful combination of coverage, resolution and sensitivity (Parker \& Phillipps 2003).
A systematic visual search of all the survey material together with
follow-up spectroscopy of identified PN candidates is now essentially
complete and has yielded $\sim1000$ confirmed new PNe. This work, known as the
Macquarie-AAO-Strasbourg H$\alpha$ PN project or ``MASH" (Parker et al. 2004 and in preparation),
has effectively doubled the known population of Galactic PNe as recorded by Acker et al.
(1992, 1996). Obvious candidate central stars of these PNe (CSPNs) can be seen only in $\leq$10\%
of these nebulae when the original H$\alpha$ and $R$-band matching survey exposures are examined.
A systematic search for the hot, blue CSPNs, based on the on-line SuperCOSMOS $B$-band images
(Hambly et al. 2001), is under way and should significantly improve this statistic.
Planned deep $U$-band imaging would aid further and permit photometric and
spectroscopic follow-up of many more candidate CSPNs.  Nevertheless,
during the 6-year spectro-

\newpage
\onecolumn
\begin{figure}
\vspace{20.0cm}
\includegraphics{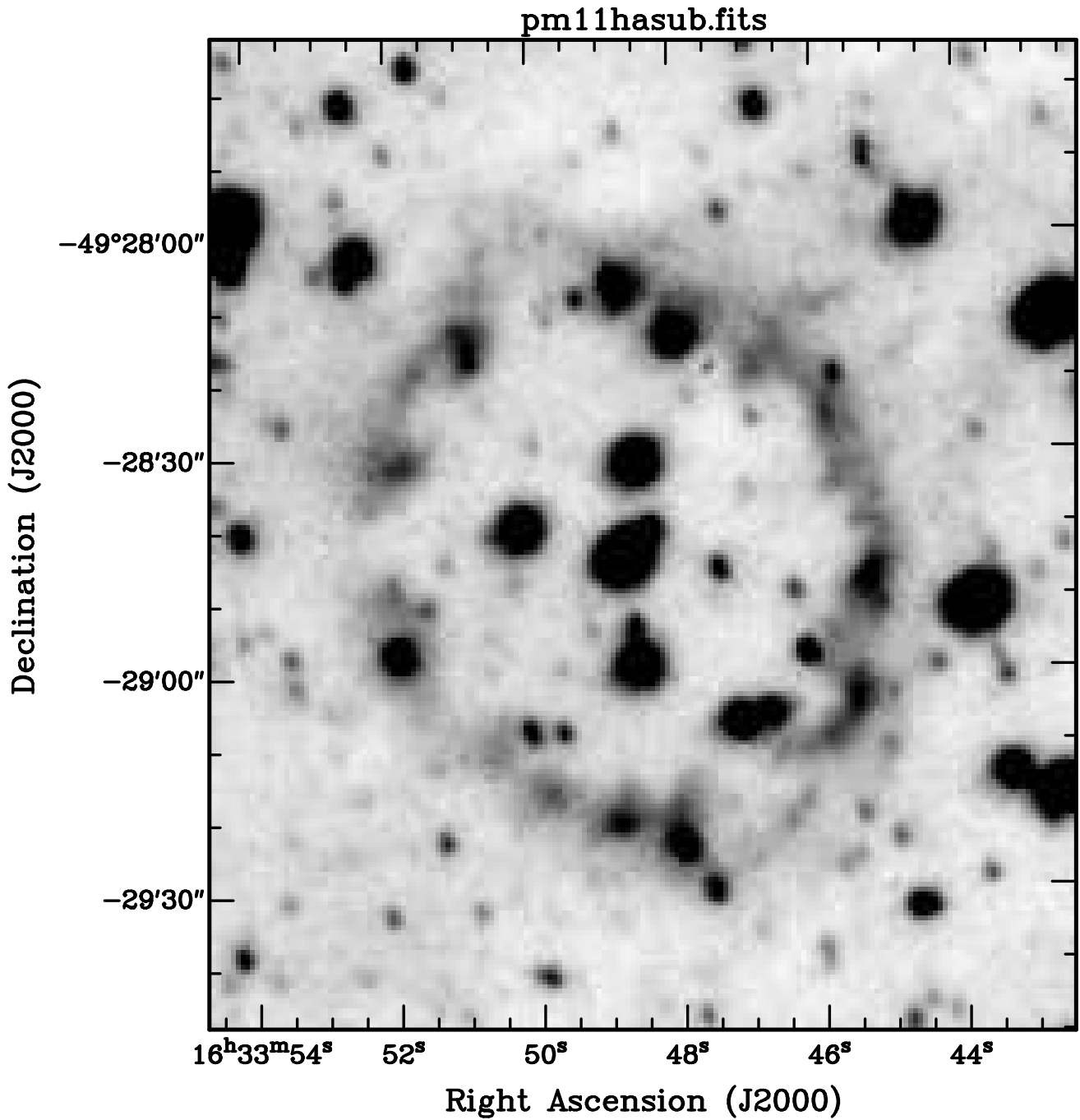}
\caption{The image of the ring of PHR1633-4928 in H$\alpha$, taken from the AAO/UKST
H$\alpha$ survey. Note the scalloping on the inside of the ring with fingers of material
extending toward the central star. The image has dimensions
118$^{\prime\prime}$$\times$135$^{\prime\prime}$.  A circle of radius 36$^{\prime\prime}$
inscribed about the central star is a good representation of the nebular rim.
The image resolution is $\sim1.5^{\prime\prime}$, with 0.67$^{\prime\prime}$ pixel size, 
and it is not flux calibrated.}
\label{haring}
\end{figure}

\twocolumn
\noindent
scopic follow-up programme, we have also discovered
seventeen PNe with central stars that show Wolf-Rayet [WR] emission features.
This represents a 30\% increase in the previous 56 known such objects listed by Jeffery et al. (1996). 
A further 14 [WR] CSPN have also been recently reported by G\'orny et al. (2005) for PN in the
Galactic Bulge.

Preliminary details for the first group of the new [WR] CSPN from MASH are documented in a
series of papers: Morgan, Parker \& Russeil (2001: ``PMR\,1,2"); Parker \& Morgan (2003:
``PM\,3,4,6,7"); Morgan, Parker \& Cohen (2003: ``PM\,5"). Several other papers are in preparation
dealing with most of the remaining sample discovered so far.
Almost all the stars are of the late [WC] or early [WC]/[WO] types with no confirmed intermediates. 
Furthermore, we have discovered PM\,5, the sole, {\it bona fide}, Galactic PN with a WN central star.    
The only other confirmed WN CSPN is N66 in the  Large Magellanic Cloud (e.g. Pena, 2000).
Unlike the previously published [WR] CPSN, many of our new [WR] CSPNs are frequently found to be 
nitrogen enriched, hinting that some subtle selection effect might be operating.
In the present paper we describe a ring nebula of rather unusual structure that was discovered
by one of us (QAP) during the search for PNe and originally catalogued as
a new PN candidate. Interestingly its central star is also of the nitrogen sequence.

At the nebular location,
Vega et al. (1980) find a weak unresolved object (VRMF~61) found from a photographic, 
objective-prism, H$\alpha$ survey.  This again neatly indicates the ability of the AAO/UKST 
H$\alpha$ survey to uncover new low surface brightness nebulosities
due to its unrivalled combination of sensitivity, resolution and coverage.
A 60-$\mu$m IRAS source (16300-4922) is also listed, with no detections in
the other three bands.  The central star of the ring is a WR
star that we believe is a Population~I object rather than a CSPN.  In \S2  we discuss the
morphology of the nebula; in \S3 the optical spectra of nebula and star; \S4 radio continuum imagery
from Molonglo; \S5, mid-infrared (MIR) MSX images; \S6, 2MASS near-infrared
(NIR) images; \S7, optical and IR photometry; and in \S8 the distance estimates to the nebula and the
stellar spectral energy distribution (SED). Our discussion is given in \S9 and
covers the nature of the nebula and the physical processes that have shaped it.

\begin{table}
\begin{center}
\caption{Details of the Ring Nebula, PCG\,11}
\label{bdata}
\begin{tabular}{lc}
\hline
{\it MASH data}& \\
Nebular centre RA (J2000)             & $16^{\rmn h}~33^{\rmn m}~48.6^{\rmn s}$  \\ 
Nebular centre Dec (J2000)            & $-49^\circ~28'~43''$                   \\   
{\it l,b}                   & 335.35$^\circ$, $-$1.14$^\circ$            \\ 
Star RA (J2000)                       & $16^{\rmn h}~33^{\rmn m}~48.74^{\rmn s}$  \\
Star Dec (J2000)                      & $-49^\circ~28'~43.5''$\\
H$\alpha$ Survey Field & HA349           \\
H$\alpha$ Survey Film  & HA18885            \\
Short Red (SR) Survey Film         & SR18884                                  \\
Date of survey images & 2000 July 8     \\
PHR designation     & PHR1633-4928                          \\
Outer dimensions (arcsec)     & 91 $\times$ 77             \\
{\it Literature}& \\
{\it IRAS} Source      & 16300$-$4922                               \\
2MASS source number& J16334874-4928439\\
{\it Parameters derived in this paper}& \\
Best fitting circle radius& 36$\pm$3\,arcsec\\
Nebular radial velocity (LSR) & $-$49~km~s$^{-1}$ \\
Distance               & 4.1$\pm$0.4~kpc \\
Radius of shell& 0.71~pc\\
Characteristic RT scale of ring & 0.24$\pm$0.02~pc\\ 
843-MHz flux density of ring& 47~mJy\\
\hline
\end{tabular}
\end{center}
\end{table}

\section{The nebula}
PCG\,11, or PHR1633-4928, was identified as a candidate PN during the MASH visual scanning phase,
and targeted as a high priority for spectroscopic follow-up due to the unusual morphology of
the nebular ring, as illustrated in Figure~\ref{haring}.  Table~\ref{bdata} presents parameters for
this nebula and its exciting star, separated into results drawn
from the MASH survey, from the literature, and derived in this paper.
The overall outer dimensions of the nebulae are 91$^{\prime\prime}$ (major axis) by
77$^{\prime\prime}$ (minor axis).
The first impression is that the nebula is oval, due primarily to the ``blow-outs" to the N-E and S-W of
the nebulae lending it a lemon-like overall outer shape.  In fact, if these two opposing regions are
ignored, its periphery is very closely circular with a radius of 36$^{\prime\prime}$,
centred on the coordinates given at the top of Table~\ref{bdata}.
We identify the exciting star as the bright SE component of the central blend of two stellar
images because of its location at the geometric centre of this circle and its WR spectrum (\S3).
Ansae extend beyond the ring to the north and south where the rim is punctured by a pair of symmetric
blow-outs along a diameter through the central star in position angle (PA) 25$^\circ$-205$^\circ$.
A third gap ocurs to the east. Arcuate filaments are seen outside the northwest rim where there is
no gap in the ring, but these may be merely part of the general, faint, diffuse background of
H$\alpha$ emission that extends to the NW for 2~arcmin.  The nebular structure is unique when 
compared with all the extant MASH PNe. A remarkable feature of the
ring is the presence of a series of scallops around the inner rim. These have a
characteristic angular scale of 12$\pm$1$^{\prime\prime}$ (the weighted average
of 18 individual features).  These strongly resemble textbook
examples of Rayleigh-Taylor (RT) instabilities.  Several of these fingers of gas
are more obvious than the nebulous rim in general, but are brightened by
foreground or background stars (\S6).  Within the ring there is even less emission (appearing
``lighter'' in this negative image) than the surrounding 
interstellar medium (ISM).  The dark cloud ``DC~328'', from the catalogue by 
Feitzinger \& Stuwe (1984), is centred at {\it l}~=~335.37$^\circ$, {\it b}~=~$-$1.14$^\circ$,
very close to the centre of PCG\,11, with an area of 1.38~deg$^2$.
A powerful stellar wind could have evacuated the inner volume of this presumably ellipsoidal nebula, reducing the projected H$\alpha$ emission except at the limb-brightened rim.
Alternatively, the dark nebular interior may indicate that its inner volume suffers heavy
local extinction due to ISM dust swept up by the stellar wind, or perhaps formed within the wind.
If so, a large Balmer decrement might be expected in any internal emission detected.

\section{Optical spectroscopy}
During the programme of spectroscopic follow-up of MASH PN candidates, long-slit red spectra of
the rim and central star of PCG\,11 were obtained on the Mount Stromlo \& Siding Springs Observatory
ANU 2.3~m telescope on 2002 July 8 using the double beam spectrograph with G1200R in the red arm.
The spectral range was 6183-6799\AA\,, the exposure was 600s, and the dispersion 0.55\AA~pixel$^{-1}$.  
Instrumental resolution was the FWHM of an arc line, $\approx$1.8 pixels, or 1\AA.  The slit
width was $\sim$2$^{\prime\prime}$. Wavelength calibration was via a standard Copper-Argon lamp. 
The spectrum was flux calibrated using observations of LT7987, a spectrophotometric standard
star from the list of Stone \& Baldwin (1983).  Additional spectra were taken on the South African
Astronomical Observatory 1.9m telescope, with a 2$^\prime$ long slit, through both
the ring and its central star, on 2002 July 17.  The spectral range was 3220-7232\AA\, at
a lower dispersion of 2.3\AA~pixel$^{-1}$ with grating G300B, and a 
resolution of $\approx$5\AA.  The exposure
was 600s. Wavelength calibration was again via Copper-Argon arc lamp exposures, and flux
calibration was achieved through observations of the same standard star LTT6248 taken
15~minutes prior to the target exposure. The slit was set wide ($\sim2.4^{\prime\prime}$) 
to allow more flux onto the detector for the nebula observations. The reported seeing was 
1$^{\prime\prime}$ so most of the flux of the central star should have passed through the 
slit. The spectra were reduced using standard {\sc IRAF} routines.

\subsection{The nebula}
\begin{figure}
\vspace{7.0cm}
\includegraphics{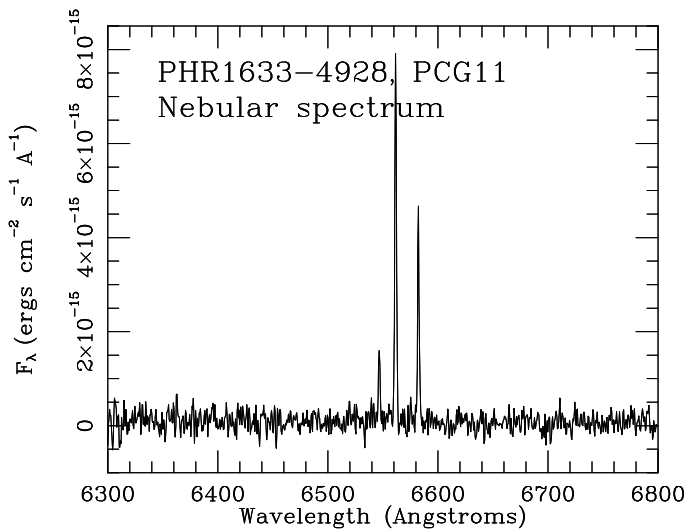}
\caption{The H$\alpha$ and [N{\sc ii}] emission lines in the ring of PCG\,11, from our single
high-resolution observation of the nebular rim.  Note the absence of the [S{\sc ii}] lines.}
\label{nebspec}
\end{figure}

Nebular emission lines from H$\alpha$ and the $\lambda\lambda6548,6584$  [N{\sc ii}] lines were
detected in three locations around the optical shell but with no hint of other common nebular
lines nor of a continuum over the wavelength range covered.
The spectrum in Figure~\ref{nebspec} obtained with the ANU 2.3m has our best signal-to-noise
ratio (SNR) and spectral resolution for the nebula.  Barycentric radial
velocities were obtained for the three strong lines as follows: H$\alpha$, $-$36; N1, $-$43;
and N2, $-$44~km~s$^{-1}$. The weighted mean velocity, corrected to the LSR, is
$-$49$\pm$5~km~s$^{-1}$. Assuming that PCG\,11's radial velocity is due solely to Galactic
rotation, one derives a distance of 3.5$\pm$0.3~kpc (Fich, Blitz \& Stark 1989; their assumed 
solar Galactocentric distance is 8.5~kpc).  The absence of the red [S{\sc ii}]
doublet argues against an identification as a PN although it would be consistent
with WR ring nebulae.  In their photographic survey of WR nebulae,
Heckathorn, Bruhweiler \& Gull (1982) found [S{\sc ii}] lines either undetected or very
faint in 8 of 10 objects associated with single stars of types WN5-8.

\subsection{The central star}
\begin{figure}
\vspace{7.0cm}
\includegraphics{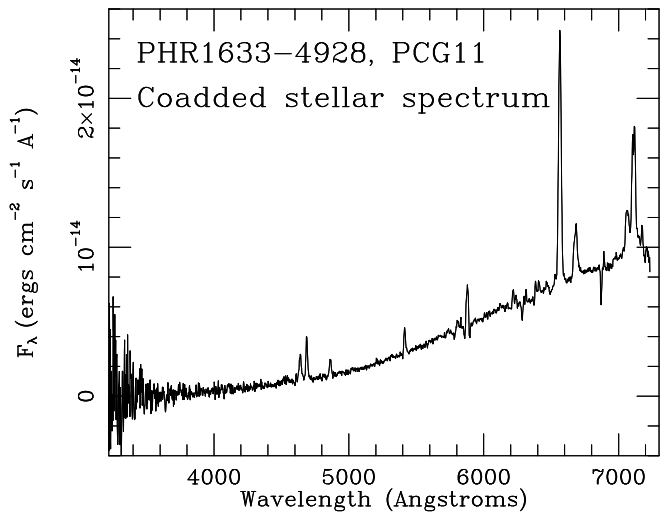}
\caption{Coadd of two low-resolution spectra of the central star in PCG\,11 over our 
complete wavelength range.}
\label{nebstar}
\end{figure}

\begin{figure}
\vspace{7.0cm}
\includegraphics{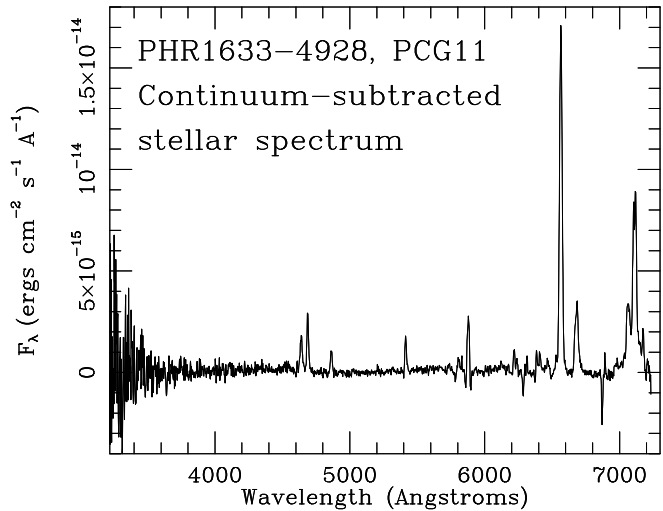}
\caption{Continuum-subtracted version of the spectrum in Fig.~\ref{nebstar}.}
\label{rectstar}
\end{figure}

The two independent lower-resolution stellar spectra of the central star of PCG\,11 taken at SAAO 
were co-added. Figure~\ref{nebstar} presents the resulting flux-calibrated spectrum over the 
complete spectral range, while Figure~\ref{rectstar} offers a continuum-subtracted version
in which the lines are more readily compared.  These spectra are perfectly adequate for line 
identification but not for the derivation of accurate radial velocities.  Our ANU 2.3~m 
high-resolution spectrum can provide good radial velocities but many of the lines included 
in its wavelength range are blends.  
Table~\ref{lines} summarizes line wavelengths, equivalent widths, and identifications.

We have firmly classified PCG\,11 as WN7h using three different approaches.  van der Hucht 
(2001, Table~2) lists descriptive criteria for line strengths, namely that:
the 4640 N{\sc iii} blend $>$ 4605,4621 N{\sc v}; 4640 N{\sc iii} $<$ He{\sc ii} 4686; and there is
a weak P Cyg profile to the 5876 He{\sc i} line.  Smith, Shara \& Moffat (1996, Table~4a)
describe their 3-dimensional classification scheme for WN stars, in which
the ratio of 5411 He{\sc ii} to 5875 He{\sc i} is the ``primary discriminant of ionization
subclass".  They quantify a series of secondary criteria, relying upon the simple ratio
of line-peak-above-continuum to continuum as the most useful measure of line strength,
and tabulating ratios of these measures for pairs of lines against spectral subtype.
For PCG\,11: 5411/5875 is 0.78 (WN7); 5808/5875 is 0.27 (WN7); and 5808/5411 is 0.33
(WN7 or 8); 4057 of N{\sc iv} is too weak to measure; and 4605 of N{\sc v} cannot be
separated adequately from 4621 in our spectrum.  All these ratios indicate a WN7 star.
Direct comparisons have been made of PCG\,11's spectrum with the spectra of
Population~I WR stars presented both by Smith et al. (1996) (as far as 6000\AA) and by
Vreux et al. (1989) at longer wavelengths.  We have focused on the appearance and
relative line strengths of the 5801-12 C{\sc iv} and 5876 He{\sc i} features, and of the
trio of the 7065 He{\sc i}, 7103-28 N{\sc iv}, and He{\sc ii} 7177 lines.  All
unequivocally suggest that PCG\,11's exciting star is most similar to a WN7.  On the 
basis of the presence of H in the spectrum (according to the criteria of Smith et al. 
(1996), who compare the heights of lines due to H+He with a line connecting pure HeII 
peaks), we assign the type WN7h.

Several unblended lines show P~Cyg absorption features from the WR stellar wind, namely 
5876\AA\ He{\sc i}, 5411\AA\ He{\sc ii}, and 5801\AA\ C{\sc iv} lines.  However, only
the He{\sc i} line represents the true wind velocity (the other lines are formed in the
inner WR wind).  From our high-resolution red spectrum we have measured the velocity for the 
5876\AA\ line, {\sc v}$_{edge}$, defined by Prinja, Barlow \& Howarth (1990) as the 
transition from stellar continuum to the beginning of the P~Cyg absorption.  This is 
$-1350\pm110~km~s^{-1}$~LSR), so the central star of PCG\,11 has 
{\sc v}$_{\infty}\leq1350~km~s^{-1}$.  Van der Hucht (2001, Table~15) compiles
estimates of terminal velocity for all the Galactic WR stars in the Seventh Catalogue.  The
average terminal velocity for the six WN7 stars in his table, without spectroscopic evidence 
of binarity, is 1300$\pm$100~km~s$^{-1}$ (standard error of the mean).  
Therefore, PCG\,11's wind velocity is consistent with a Population~I WN7 star.  

\begin{table}
\begin{center}
\caption{Emission lines observed in the central star of PCG\,11.}
\label{lines}
\begin{tabular}{lrl}
\hline
$\lambda$ observed& EW& Identification\\
(\AA)& (\AA)&        Ion, air wavelength (\AA)\\
4610&       $-$2.9&      N{\sc v} 4605.0,4621.3\\
4634-41&    $-$36&      N{\sc iii} 4634.2,4640.6,4641.2\\
4686&       $-$42&      He{\sc ii} 4687.0\\
4864 &  $-$15&     N{\sc iii} 4858.7,4867.2; He{\sc ii} 4859.3;\\
          &   &    N{\sc iv} 4867.2\\
5412&        $-$16&     He{\sc ii} 5411.2\\
5801,12&     $-$2.0&     C{\sc iv} 5801,12\\
5876&        $-$9.1&     He{\sc i} 5875.6\\
6219&        $-$3.5&     N{\sc iv} 6219\\
6239&        $-$2.0&     He{\sc ii} 6234?\\
6311&        $-$1.8&     He{\sc ii} 6310.8\\
6409&        $-$3.3&     He{\sc ii} 6406.?\\
6450-80 blend& $-$3.6&    N{\sc iii} 6450,60;+?\\
6527&     $-$1.9&     N{\sc iv} 6527.8\\  
6562&  $-$58&     He{\sc ii} 6560.1; H{\sc i} 6562.80\\
6678&       $-$13&      He{\sc i} 6678.1; N{\sc iv} 6678\\
6892&       $-$0.9&      He{\sc ii} 6890\\ 
7063&       $-$12&      He{\sc i} 7065.20\\
7100-7130&  $-$33&      N{\sc iv} 7103$-$7128\\
7177&       $-$4.5&      He{\sc ii} 7177.52\\
\hline
\end{tabular}
\end{center}
\end{table}

\noindent
\section{Molonglo radio continuum images}
\begin{figure}
\vspace{8.5cm}
\includegraphics{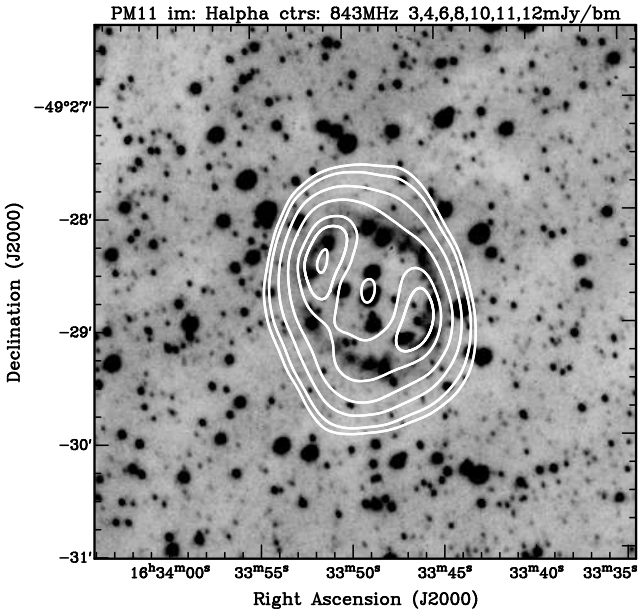}
\caption{The H$\alpha$ image of PHR1633-4928 overlaid by white (positive)
contours from the coaddition of three independent Molonglo 843-MHz
images.  The peak of the radio image is 12.1~mJy beam$^{-1}$, and the noise in this combined 
image is 1.0~mJy beam$^{-1}$.  The contours correspond
to values of 3, 4, 6, 8, 10, 11, and 12~$\sigma$. The small central contour that includes
the exciting star is at 8~mJy beam$^{-1}$ (8~$\sigma$).}
\label{hamost}
\end{figure}

Three independent images are available from the Molonglo Galactic Plane Surveys (MGPS1, MGPS2:
Green et al. 1999; Green 2002) at 843~MHz: from MGPS1 on 1989 May 15; and from MGPS2 on 2000 August 17 
and 22.  The r.m.s. noise in each image is 3.0, 1.5, and 2.0~mJy beam$^{-1}$, respectively.
We used the task {\sc imcomb} in {\sc miriad} to combine the three images using inverse-variance
weighting.  The resultant image appears in the form of
white contours overlaid on the H$\alpha$ image in Figure~\ref{hamost}.  The noise in the
combined image is 1.0~mJy beam$^{-1}$.

There are twin peaks in the radio map on the northeast and southwest rims of the H$\alpha$
ring, both with values of 12~mJy beam$^{-1}$.  
The emission falls at the middle of the image to a local minimum of 8~mJy beam$^{-1}$ at the
location of the central star.  The MGPS beam size is 43$^{\prime\prime}$ (in RA) by
56$^{\prime\prime}$ (in DEC).  Given the angular resolution, we can conclude only that the
radio continuum emission is resolved and appears to be associated with the H$\alpha$ rim with two peaks,
diametrically opposed on the ring, and likely not aligned with the brightest sections
of the H$\alpha$ ring.  The spatial integration of the combined MGPS image yields a
total flux density of 47~mJy (corrected for the local background).  The total integrated radio 
continuum flux densities of well-known, previously studied WR ring nebulae lie between 2.6 and
11~Jy (Johnson \& Hogg 1965; Wendker et al. 1975; Cappa, Goss \& Pineault 2002).  If we
compute 1.4-GHz luminosities for WR ring nebulae (4\,$\pi$\,distance$^2$\,S(1.4-GHz))
from the literature these average $4\pm1.6\,\times10^{15}$~W~Hz$^{-1}$ (NGC~6888, 1.26~kpc; 
NGC~2359, 3.67~kpc; WR~101, 3.18~kpc; WR~113, 1.79~kpc; all distances are taken from 
van der Hucht (2001, Table~28)).  The corresponding figure for PCG\,11 is 
$1.0\times10^{14}$~W~Hz$^{-1}$, some 40 times smaller.  This suggests that PCG\,11 is more 
distant than the 1.3-3.7~kpc of these previously known Galactic WR ring nebulae, 
and/or that the mass loss rate of its exciting star is substantially below those of WR stars
associated with such nebulae.  Unfortunately, the NRAO VLA Sky Survey (NVSS) has a southern declination limit of 
$-$40$^\circ$ while the Southern Galactic Plane Survey (SGPS) has a Galactic latitude limit of 
$-$1.0$^\circ$ for its continuum data.  In the Parkes-MIT-NRAO (PMN) Surveys, PCG\,11
can be identified as a local peak on a ridge of bright emission in PMN
4850-MHz images but the source is so weak against this confusion that it was excluded
from the point source catalogue of the Southern Survey (declination
$-87^\circ$ to $-37^\circ$).  

\section{Mid-infrared images from the Midcourse Space eXperiment (MSX)}
\begin{figure}
\vspace{8.0cm}
\includegraphics{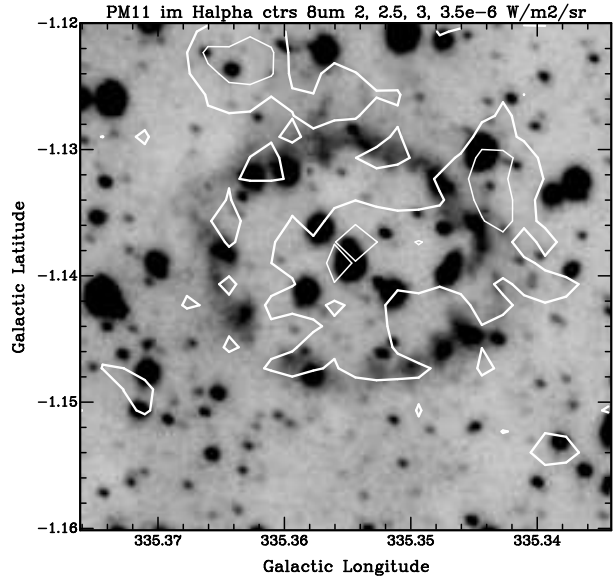}
\caption{The H$\alpha$ image of PHR1633-4928 is overlaid by white contours of MSX 8.3-$\mu$m
emission with values of 2, 2.5, 3, and 3.5$\times$10$^{-6}$~W~m$^{-2}$~sr$^{-1}$.
These correspond to 6, 7.5, 9, 10.5~$\sigma$ detection levels based on the noise in the 8.3-$\mu$m 
image.}
\label{hamsx8}
\end{figure}

\begin{figure}
\vspace{8.0cm}
\includegraphics{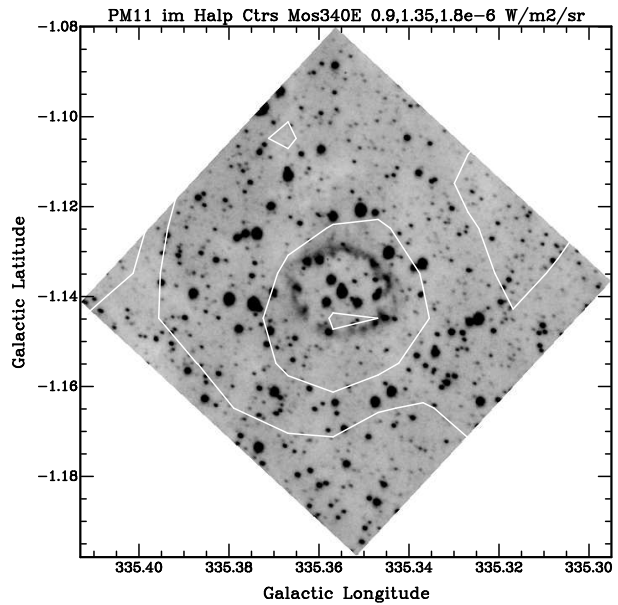}
\caption{The H$\alpha$ image of PHR1633-4928 is overlaid by white contours of MSX 21.3-$\mu$m
emission from a deep mosaic, with values of 0.9, 1.35, 1.8$\times$10$^{-6}$~W~m$^{-2}$~sr$^{-1}$.
These correspond to 5, 7.5 and 9~$\sigma$ detection levels based on the noise in
the 21.3-$\mu$m mosaic. The image is roughly 6$^{\prime}$$\times$6$^{\prime}$.}
\label{hamos21}
\end{figure}

The MSX images of the Galactic Plane with 6$^{\prime\prime}$ pixels and 20$^{\prime\prime}$
resolution (Price et al. 2001) show detections of the central star at 8.3~$\mu$m, and of the
adjacent star (seen in the H$\alpha$ image: Figure~\ref{haring}) at 8.3 and 12.1~$\mu$m.
Weak diffuse MSX 8.3-$\mu$m and $IRAS$ 12-$\mu$m emission are found in this area and most of the
interior and rim of the nebula are detected at 8.3~$\mu$m (Figure~\ref{hamsx8}).  At 
21.3~$\mu$m four diffuse patches are detected around the nebular rim, all at the 4$\sigma$ level.
Examination of the low-resolution MSX mosaic products with
36$^{\prime\prime}$ pixels and 72$^{\prime\prime}$ resolution discloses diffuse emission
covering PCG\,11's whole ring at 8.3, 14.6, and 21.3~$\mu$m. The 21.3-$\mu$m mosaic image 
(Figure~\ref{hamos21}) essentially mimics the $IRAS$ Sky Survey Atlas (ISSA) image at 
25~$\mu$m, but at higher spatial resolution.

Spatial integrations of the 8.3-$\mu$m emission associated with PCG\,11's ring
indicate 0.8$\pm$0.1~Jy (Fig.~\ref{hamsx8}) and 0.85$\pm$0.2~mJy (from the MSX deep
mosaic image), each corrected for local background emission.

\section{2MASS images}
\begin{figure}
\vspace{8.0cm}
\includegraphics{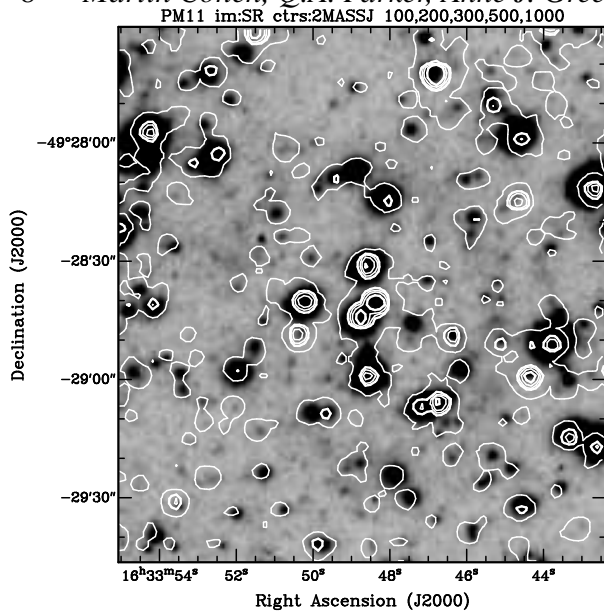}
\caption{The 2MASS $J$ image of the region of PHR1633-4928 is overlaid in white contours
on the SR image.  $J$-band contours correspond to values of 100, 200, 300, 500, and
1000 counts.  The image is approximately 2$^{\prime}$$\times$2$^{\prime}$.}
\label{2mjsr}
\end{figure}

The 2MASS images in $J$, $H$, and $K_s$ were extracted and overlaid on both the H$\alpha$ image
and the matching Short Red (SR) exposure.  There are no signs of diffuse NIR
emission associated with the rim of the H$\alpha$ oval.  Hints of the nebulous ring
are seen in the SR image where it is easier to distinguish faint stars from H$\alpha$
nebulosities.  Of particular interest is whether
any of the Rayleigh-Taylor-like fingers of H$\alpha$ nebulosity emit in the NIR, but the
low latitude of PCG\,11 means there are many field stars projected against the rim.
We have superimposed the 2MASS $J$ image as white contours on the SR
exposure of PCG\,11 (Figure~\ref{2mjsr}).
All the seeming NIR detections of the nebular rim can be identified with faint stars
in the SR image (the same is true for the $H$ and $K_s$ band images).  
We conclude that none of the H$\alpha$ fingers of gas is
detected by 2MASS, eliminating the possibility of significant shock emission in the
[Fe{\sc ii}] 1.64-$\mu$m in the $J$-band, and both the Br$\gamma$ and H$_2$ S1~1-0
2.12-$\mu$m lines in the $K_s$-band.

\section{Optical and infrared photometry}
SuperCOSMOS photometry is available on-line for PCG\,11 from digitizations of the broadband
UKST $B_J$, $R$, $I$, ESO $R$, and the H$\alpha$ and associated SR images.  The faint
companion is not deblended into a separate object in any of these photometric parameterizations.
Rather similar photometry can be formed from magnitudes for PCG\,11 from the USNO-B1.0 archives
(Monet et al. 2003: the star identifier is 0405-0535599) although we prefer the
SuperCOSMOS data because of its direct traceability to CCD calibrations from field to field,
that assures consistency of photometry.  Vega et al. (1980) report a $V$
magnitude from a photographic plate which surely refers to both stars because these
would have been unresolved at their plate scale.  The SuperCOSMOS $I$ is considered unreliable 
(the IV-N emulsion used for this band is hypersensitised with liquid silver nitrate solution often
resulting in gross background variations) as the companion brightens with wavelength, while the 
SuperCOSMOS $B_J$
will be most accurate as the companion has a negligible influence on the blend.  For the
$R$ magnitudes, the faint companion is $\sim18^m$, and affects the photometry very little,
but the strong red emission lines contaminate the bandpasses.  Consequently, we are left
solely with the UKST $B_J$ point to compare with the calibrated SED.  
This point lies approximately 15\% below the spectrum, or 0.15$^m$,
equal to the $1.5\times$ the expected 1~$\sigma$ uncertainty on the UKST photographic plate.  
This precludes any substantial miscalibration of the SED of the central star of PCG\,11.
We plan to pursue multi-colour CCD photometry at higher angular resolution for this source 
to refine the photometry and demarcation between the close stellar companions at the centre.

Table~\ref{tphot} summarizes our SuperCOSMOS 
$B_J$ photometry for the central star of PCG\,11
together with 2MASS $JHK_s$, and the MSX detections obtained by aperture photometry from
the images.  There is excellent agreement between our optically determined positions for
both the central star and its companion and those measured by 2MASS.  The average position 
for the central star is (J2000) $16^{\rmn h}~33^{\rmn m}~48.74^{\rmn s}$ $-49^\circ~28'~43.5''$
(Table~\ref{bdata}), and for the companion is 
$16^{\rmn h}~33^{\rmn m}~48.36^{\rmn s}$ $-49^\circ~28'~40.1''$.

Thus, the star adjacent to the exciting star of PCG\,11 is both redder and brighter
in the NIR and MIR than the exciting star itself.  It is even detected by MSX at
12.1~$\mu$m as well as at 8.3~$\mu$m, while PCG\,11's central star
is not.  Nonetheless, the companion lies 5.4$^{\prime\prime}$ away from the WR
star and is offset from the ring's centre, making it unlikely to be responsible
for the nebula.  One can match this star's NIR colors and magnitudes to either an
M-giant or an O-rich AGB star, with an estimated distance of 2.4~kpc and extinction,
A$_V$ of 5.8$\pm$0.4.  With these properties it represents a chance projection
along our line-of-sight to the WR star.

The $IRAS$ Point Source Catalog
contains a 60-$\mu$m-only object at the location of PCG\,11.  $IRAS$ does detect emission
from the relevant ISSA field,  but securely only at 25 and 60~$\mu$m. This information appears
in Table~\ref{tphot}, in the central star section, although the companion could also contribute
to this FIR emission if it possessed a cool, detached, circumstellar dust envelope.

\section{Distance to PCG\,11}
\begin{figure}
\vspace{8.0cm}
\includegraphics{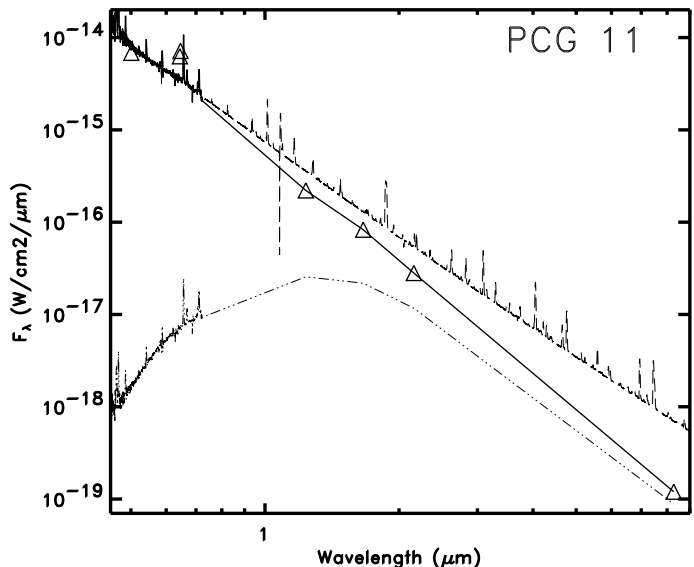}
\caption{The observed (dash-dotted) and 8.1$^m$ dereddened (solid) energy distributions of PCG\,11's WR
star, together with dereddened optical, 2MASS, and MSX photometry (triangles), 
compared with a scaled version of the modeled SED for HD~151932 (WN7h) (long-dashed).}
\label{sed}
\end{figure}

The stellar coadded spectrum is validated by the $B_J$ magnitude; i.e. essentially
all the stellar flux passed into the 2.4$^{\prime\prime}$ slit.  Synthetic photometry derived from the spectrum 
gives $b$=17.72, $v$=15.80, in the nomenclature of the narrowband, line-free, optical magnitudes 
devised by Westerlund (1966) and Smith (1968a,b).  The observed $b-v$ of 1.92
together with an intrinsic $b-v$ index of $-$0.18 (van der Hucht 2001),
indicates an extinction, A$_{\it v}=8.9$ (equivalent to A$_V=8.3$). A more precise
estimate of A$_V$ comes from dereddening our spectrum of PCG\,11's central
star to match the optical slope of the modeled SED for a single WN7h star (HD~151932), 
kindly provided by Crowther(2005).  We settled on the range 4500-7200\AA\ to avoid the high
noise at the blue end of our spectrum) and carried out formal $\chi^2$ fitting for a variety
of different extinctions.  This suggests A$_{\it v}$~=~$8.7\pm0.2$ or A$_V$~=$8.1\pm0.15$ 
(see Figure~\ref{sed}).  We adopt these values as more robust than depending on synthesized
$b$ and $v$, and on an average intrinsic $b-v$.  There is no difference between the mean 
M$_{\it v}$ of 3 binary WN7h stars ($-6.2\pm0.4$) and the only single WN7h star ($-6.56$) 
(van der Hucht 2001, Table 25) which are associated with OB associations at known distances.  
Therefore, we take M$_{\it v}$=$-6.3\pm0.3$ for all WN7h stars.  This yields a 
photometric distance to PCG\,11 of $4.7^{+1.0}_{-0.8}$~kpc.  Using inverse-variance weighting
we combine the photometric distance estimate with that based on the nebular radial velocity
(\S3.1) to obtain $4.1\pm0.4$~kpc, which we adopt hereafter.

Out to distances of 2 and 3~kpc from the sun, in the Galactic direction of PCG\,11 in the
plane, one expects a total interstellar extinction of $\geq$3$^m$ (Fitzgerald (1968: Fig.~3(j), 
zone 68); Lucke (1978: Fig.~6)).  
From the upper limit to the H$\beta$ line in our high-resolution 
nebular spectra ($<1\times10^{-15}$\,erg\,cm$^{-2}$\,s$^{-1}$) we obtain an
intensity ratio of H$\alpha$/H$\beta$ $>18$, compared with Case-B recombination
at T$_e$$\sim$10$^4$\,K of 2.86, so the total extinction to PCG\,11's rim is $>5.5\pm0.4^m$.  
Using the extinction maps of Schlegel, Finkbeiner, \& Davis (1998), the total Galactic
extinction along the line-of-sight through PCG\,11 is A$_V$~$\sim$12$^m$. Distributing this 
evenly along the line-of-sight suggests that at least 3$^m$ occurs in front of PCG\,11.  
The most direct estimate of the line-of-sight nebular extinction comes from comparing 
the predicted and observed
H$\alpha$ fluxes from PCG\,11.  The thermal radio emission provides an estimate of the intrinsic 
emission line flux in H$\alpha$ (see Condon (1992), eqns. (3) and (4a)).  Using this
formulation predicts 3.7$\times10^{-11}$~erg~cm$^{-2}$~s$^{-1}$ for the integrated
intrinsic H$\alpha$ flux, before the effects of extinction. We estimate the observed, spatially 
integrated H$\alpha$ line flux of PCG\,11 from the absolutely calibrated, but spatially much
lower-resolution, Southern H$\alpha$ Sky Survey Atlas (SHASSA: Gaustad et al. 2001).  At PCG\,'s
location this survey shows 122~Rayleighs, equivalent to 
$1.7\times$10$^{-12}$\,erg\,cm$^{-2}$\,s$^{-1}$).  
The SHASSA H$\alpha$ filter also includes the red [N{\sc ii}] lines.  We have corrected for
these based on the profile of the SHASSA filter shown by Gaustad et al. (2001) and on our
spectrum of PCG\,11's nebula (Figure~\ref{nebspec}).  The corrected SHASSA H$\alpha$ line flux is
then $1.2\times$10$^{-12}$\,erg\,cm$^{-2}$\,s$^{-1}$.  The ratio of observed to predicted line 
fluxes indicates A$_V$~=~4.6$^{+0.4}_{-0.2}$ magnitudes, including estimated uncertainties.  
We adopt this as the most reliable estimate of the line-of-sight extinction between the Sun and 
PCG\,11.  Comparing with the stellar reddening nebular Balmer decrement (above) we find
unaccounted for extinctions of $\sim3\pm0.5^m$ toward the central star, and $1\pm0.5^m$ toward 
the nebular rim.  The roughly $2\pm0.7^m$ we attribute to the shell of swept up ISM 
surrounding PCG\,11.  This would explain why the interior of the nebula has so little 
emission compared with the environs of PCG\,11 (Fig.~\ref{haring}).

The resulting dereddened spectrum follows the Crowther model as far as the limit of our
optical spectra (Fig.~\ref{sed}), after which it declines more rapidly than the model SED.  
PCG\,11's smaller IR free-free emission implies a lower mass loss rate than that of
HD~151932 which is 6.6$\times$10$^{-5}$~M$_\odot$~yr$^{-1}$, assuming a
homogeneous shell (Crowther 2005).  Typical values for WN7 stars are 
$3-6\times10^{-5}$~M$_\odot$~yr$^{-1}$ (Leitherer, Chapman \& 
Koribalski 1995; Prinja et al. 1990).  A rough estimate can be made for the mass loss
rate of PCG\,11's star based on the excess 8.3-$\mu$m flux density above the stellar
photosphere.  Allowing for a Rayleigh-Jeans contribution of the hot photosphere at 
8.3~$\mu$m, the excess, dereddened (for A$_V$=8.1$^m$), flux density at this wavelength 
is $\sim$15~mJy.  We assume that this emission is entirely due to free-free emission.
The mass loss rate (Wright \& Barlow 1975) would be 
$\sim1.0\pm0.3\times10^{-5}$\,M$_{\odot}$\,yr$^{-1}$
(with MIR Gaunt factor from Beckert et al. (2000), and mean molecular weight, ionic charge, and
number of electrons per ion for a WN7 star from Barlow, Smith \& Willis (1981)).
The error is dominated by the uncertainty in the photospheric level, with a component
from the error in extinction correction.  No corresponding estimate of stellar mass loss 
is possible from the 843-MHz radio continuum because this is resolved and comes from the 
nebula at large, not just the central star.  

\section{Discussion}
The emission-line spectrum of PCG\,11 marks it as a WN7h star.  At a distance of 4.1~kpc
it lies 80~pc below the plane.  This is in accord with the distribution of Galactic
WR stars (van der Hucht 2001, Fig.~8), their mean absolute z-distance (49~pc: van der 
Hucht 2001), and the 45-pc scale height of WRs (Conti \& Vacca (1990) and of OB stars 
(Garmany et al. 1982).  From the N-richness of PCG\,11's central star and this apparent kinship 
with massive stars we conclude that PCG\,11 is not a PN but rather a Population~I WR ring 
nebula.  Are other properties of this nebula in accord with this classification?

Two classical ring nebulae associated with Population~I WN stars are NGC~2359 (HD~56925: WN4)
and NGC~6888 (HD~192163: WN6(h)), and both were first studied in the radio continuum by
Johnson \& Hogg (1965).  Their single-dish measurements indicated a thermal radio spectrum.
This result was confirmed by Wendker et al. (1975) using the Effelsberg 100-m dish and
aperture synthesis at three frequencies.  Wendker et al.
also commented on the close agreement between the H$\alpha$ and radio continuum
morphologies.  Furthermore, thermal spectra were derived for two ring nebulae around WC8 stars
(WR~101 and 113) by Cappa et al. (2002), who compared spatially integrated 1.465-GHz VLA
flux densities with those from the 4.85-GHz survey by Haynes, Caswell \& Simons (1978).

Although PCG\,11 has only a single radio continuum measurement (from the MGPS), 
Cohen \& Green (2001) have calibrated the ratio of the spatially integrated 8.3-$\mu$m
and MGPS 843-MHz flux densities for a wide range of thermal and nonthermal
structures in their study of the ISM near {\it l}=312$^\circ$.  These authors found
a characteristic ratio of flux densities with a median of 24 for thermal sources,
and $\leq$3 for nonthermal.  We have calculated
this ratio for PCG\,11's ring using both the 6$^{\prime\prime}$ pixellated MSX
image and the coarser resolution but deeper mosaic image at 8.3~$\mu$m against the
843-MHz flux.  The ratio is 19, confirming that this emission is thermal in nature.

Spatial integration of the identical area of the ring from the 8.3, 14.6, and
21.3-$\mu$m mosaic images yields flux densities at three wavelengths of 0.85,
5.5, and 11.0~Jy, respectively.  Combining these with the $IRAS$ 60-$\mu$m
flux density of 14.6~Jy (because the $IRAS$ beam is large enough to accommodate
the entire ring), leads to an SED well fitted by a 170~K
blackbody at the three longest wavelengths.  At 8.3~$\mu$m the blackbody accounts
for only about 60\% of the observed emission.  But we expect the diffuse ISM to 
contribute significant flux in the 8.3-$\mu$m bandpass due to polycylic aromatic hydrocarbon
(PAH) emissions at 6.2, 7.7 and 8.7~$\mu$m, elevating and perhaps overcorrecting for sky 
emission in our off-source locations.
Mathis et al. (1992) analyzed the IRAS emission from three WR ring nebulae.
They interpreted the low levels of 12-$\mu$m emission to imply that the
dominant mechanism was continuum radiation from dust grains, rather than PAH emission
bands.  Thermal dust emission also characterizes these nebulae at 25, 60, and 100~$\mu$m
but, to explain the brightness at 25~$\mu$m, Mathis et al. (1992) argued for transient heating of
small grains in excess of 100~K.  Our 170~K derived color temperature between 14.6, 21.3 and 
60~$\mu$m in PCG\,11 would be consistent with transiently heated small grains.  The weakness of 
the 8.3-$\mu$m emission detected in PCG\,11 (Fig.~\ref{hamsx8}) also suggests the absence of 
PAH emission, probably due to destruction by the WR star's wind.  

Following Mezger \& Henderson (1967, Appendix A, equations A.12, A.13, 
and A.14), the observed radio flux density 
can be converted into the mean density and mass of an ionized nebula. Deconvolving the 
843-MHz image (Fig.~\ref{hamost}) leads to an average full power width (FPW)
of 1.6$^\prime$.  Adopting the simplest model of a uniformly filled sphere to represent 
PCG\,11, we derive an equivalent Gaussian FPW 1.1$^\prime$ that can be used in the formulation
of Mezger \& Henderson.  For an electron temperature of 10$^4$~K, the 843-MHz
flux density indicates ionized mass of 4.2~M$_\odot$.
To inject this much ionized matter into the ISM with the mass loss rate typical of a 
WN7 star (cited above) would have taken $\sim$100,000~yr.  
McCray (1983, Table~1) summarizes the properties that characterize
WR ring nebulae as radii 0.3$-$10~pc, ages 20,000-200,000~yr, and shell masses 5-20~M$_\odot$.
PCG\,11, therefore, presents a consistent picture as a small WR ring nebula.

The total mass (neutral and ionized) of the ISM outside the H$\alpha$ shell is constrained
by the need to produce the additional extinction, local to the nebula, of $\sim2^m$ (\S8).
The column density in H{\sc i} would be $4\times10^{21}$\,H~cm$^{-2}$, above a shell
of radius 0.71~pc.  The total mass required is 200~M$_\odot$, quite acceptable for a WR
shell (e.g. Chu 1982).  It is clear that the accumulated ISM mass dominates the ejecta from
the red supergiant precursor of the WN7 star, and that only a small fraction of the
matter is ionized.  Diffuse 21.3-$\mu$m emission appears confined to the rim where it is
enhanced by limb-brightening (\ref{hamos21}).  In \S4 we derived
a dust temperature of 170~K from this emission.  Assuming that dust and gas are well-coupled
at the periphery of the shell, we can assign the same temperature to the gas, leading to
a sound speed of $\sim$0.7~km~s$^{-1}$.  

Scalloping appears around the whole interior of PCG\,11 in gas that is a combination of
swept up ISM and stellar ejecta.  This is sandwiched between shocked ambient gas and an inner contact 
discontinuity at the interface with the stellar wind.  Weaver et al. (1977) demonstrate that 
RT instability commonly develops in expanding bubbles due to the collapse of this thin boundary.  
The fastest-growing RT mode dominates the observed structure.  From McCray 
\& Kafatos (1987), but neglecting magnetic fields, the scale size of the dominant RT instability 
is given by 9$a_s$$^2$/(4{\it G}$\rho_0${\it R$_s$}), with {\it G} the gravitational constant, 
$\rho_0$ the density of the shocked ISM (1.3\,n$_0$\,M$_H$), and {\it R$_s$} the shell radius.  
With a$_s$ of 0.7~km~s$^{-1}$, a wavelength for the observed structure of 0.24~pc, and a shell
radius of 0.71~pc, the shocked ISM density (n$_0$) would be 48000\,cm$^{-3}$.  Such a 
density would explain the absence of [S{\sc ii}] lines and the presence of [N{\sc ii}] emission in 
the rim because it lies between the critical densities of 6731\AA\, of [S{\sc ii}] and 6584\AA\, 
of [N{\sc ii}].  The red [S{\sc ii}] doublet lines would be quenched.  
Allowing for the projection of the curved shell and for blurring by seeing, we estimate the 
thickness of the limb-brightened H$\alpha$-emitting rim of PCG\,11 to be $\sim2.2^{\prime\prime}$.
To accommodate the 4.2~M$_\odot$ of ionized gas within this region implies a density of
roughly 500\,H~cm$^{-3}$.  This region is at least partially ionized and for simplicity we take 
N$_i$=N$_e$.  The resulting filling factor for ionized gas would thus be 0.02, consistent with the 
range derived for the nebula around WR113 by Cappa et al. 2002).  Once instabilities are established,
their growth proceeds with a growth time of $\sim$130,000~yr, comparable to the time required to
create the bubble (see above). 

\section{Conclusions}
We characterize PCG\,11 as a wind-blown bubble generated by a well-centred 
Population~I Wolf-Rayet star of type WN7h.  The stellar mass loss rate is much smaller 
than those of other WR stars in ring nebulae, and the ring is more distant from the
Sun than these.  Both contribute to PCG\,11's very low radio luminosity.  This
is proportional to \.{M}$^{4/3}$ (Wright \& Barlow 
1975).  The factor of 5 between PCG\,11's \.{M} and the average for WN7 stars 
($5\times10^{-5}$\,M$_{\odot}$\,yr$^{-1}$) reduces radio luminosity by almost
an order of magnitude.  Further reduction may be due to the low filling factor for ionized gas.
In producing the almost spherical ionized 
volume, approximately 200~M$_\odot$ of the ISM was swept up by the advancing shell.
Material at the inner boundary of that enveloping material suffers Rayleigh-Taylor instability.
This appears as a highly regular scalloping of the inner margin of the nebulous 
H$\alpha$ ring, defined by fingers of infalling ionized matter.  
There is a very well-defined wavelength for this phenomenon in PCG\,11 and it
is observed around the entire nebular rim.  No PN has the character of PCG\,11's
shell.  The closest approach might be NGC\,6894 which shows two or three large-scale
scallops inside its bright rim.  However, the outer edge of the rim is distorted, not 
circular, where these indentations occur and its overall structure may owe more to 
stripping by the ISM than to gravitational instabilities (Soker \& Zucker 1997).
Although RT instabilities are known on larger scales (e.g. McClure-Griffiths et al. 2003),
PCG\,11 is the only example of a {\it complete} shell of these instabilities of which we 
are aware, on any observable spatial scale.

\section{Acknowledgments}
We thank Mark Wardle, Jessica Chapman, Naomi McClure-Griffiths, and David Frew for valuable
discussions.  We are grateful to the referee, Paul Crowther, for his extremely useful comments
on this paper, and for offering his model energy distribution of a WN7 star.
MC thanks NASA for supporting this work under LTSA grant, NAG5-7936, 
and ADP grant, NNG04GD43G, with UC Berkeley.  QAP acknowledges the joint support 
of the AAO and Macquarie University.  The MOST is owned and operated by the School 
of Physics, within the University of Sydney, and supported by the Australian 
Research Council and the University of Sydney.  We thank ANSTO for the provision 
of observing support to QAP to enable spectroscopic follow-up. This research made use 
of the AAO UKST H$\alpha$ survey made available to the community on-line from the Wide 
Field Astronomy Unit at the Royal Observatory Edinburgh.  This work also made use of 
data products from the Midcourse Space Experiment.  Processing of the data was funded 
by the Ballistic Missile Defense Organization with additional support from NASA 
Office of Space Science.  This research has also made use of the 
NASA/IPAC Infrared Science Archive, which is operated by the 
Jet Propulsion Laboratory, California Institute of Technology, 
under contract with the National Aeronautics and Space Administration.

\begin{table}
\begin{center}
\caption{Optical \& IR photometry of the central star of PCG\,11}
\label{tphot}
\begin{tabular}{lc}
Waveband &  Mag/Flux Density\\ 
\hline
{\it The Central Star}& \\
{\it b}[synthetic]    & 17.72 \\
{\it v}[synthetic]    & 15.80 \\
$B_j$                 & 16.3$\pm$0.2 \\
$ESO-R$                & 12.9$\pm$0.1 \\
$UKST-R$               & 13.1$\pm$0.1 \\
$J$                    & $ 10.22\pm0.03$  \\
$H$                    & $ 9.29\pm0.06$ \\
$K_s$                    & $ 8.91\pm0.07$ \\
{\it MSX} F(8.3-$\mu$m) (mag)[mJy]& 8.60 [21]   \\
\hline
{\it The companion star}& \\
$J$                    & $ 9.68\pm0.05$  \\
$H$                    & $ 8.19\pm0.04$ \\
$K_s$                    & $ 7.61\pm0.03$ \\
{\it MSX} F(8.3-$\mu$m) (mag) [mJy]            & 7.71 [46]   \\
{\it MSX} F(12.1-$\mu$m) (mag) [mJy]           & 7.84 [19]   \\
\hline
{\it PCG\,11 nebula}& \\
{\it MSX} F(8$\mu$m: 6$^{\prime\prime}$ pix) [Jy]            & $ 0.79 $  \\
{\it MSX} F(8$\mu$m: 36$^{\prime\prime}$ pix) [Jy]            & $ 0.85 $  \\
{\it MSX} F(15$\mu$m: 36$^{\prime\prime}$ pix) [Jy]            & $ 5.50 $  \\
{\it MSX} F(21$\mu$m: 36$^{\prime\prime}$ pix) [Jy]            & $ 11.0 $  \\
{\it IRAS} F(12$\mu$m) (Jy)           & $<3.0 $    \\
{\it IRAS} F(25$\mu$m) (Jy)           & $<2.4 $    \\
{\it IRAS} F(60$\mu$m) (Jy)            & $14.6\pm1.2 $    \\
{\it IRAS} F(100$\mu$m) (Jy)            & $<470 $    \\
\hline
\end{tabular}
\end{center}
\end{table}

\end{document}